\documentclass[conference]{IEEEtran}
\IEEEoverridecommandlockouts
\usepackage{cite}
\usepackage{amsmath,amssymb,amsfonts}
\usepackage{algorithmic}
\usepackage{graphicx}
\usepackage{textcomp}
\usepackage{subfigure}
\usepackage{xcolor}
\usepackage{url}
\def\BibTeX{{\rm B\kern-.05em{\sc i\kern-.025em b}\kern-.08em
    T\kern-.1667em\lower.7ex\hbox{E}\kern-.125emX}}
\begin{document}

\title{Convolutional Transformer-Based Image Compression\\

}

\author{\IEEEauthorblockN{Bouzid Arezki}
\IEEEauthorblockA{\textit{L2TI Laboratory} \\
\textit{University Sorbonne Paris Nord}\\
Villetaneuse, France \\
bouzid.arezki@edu.univ-paris13.fr}
\and
\IEEEauthorblockN{Fangchen Feng}
\IEEEauthorblockA{\textit{L2TI Laboratory} \\
\textit{University Sorbonne Paris Nord}\\
Villetaneuse, France \\
fangchen.feng@univ-paris13.fr}
\and
\IEEEauthorblockN{Anissa Mokraoui}
\IEEEauthorblockA{\textit{L2TI Laboratory} \\
\textit{University Sorbonne Paris Nord}\\
Villetaneuse, France \\
anissa.mokraoui@univ-paris13.fr}

}

\maketitle

\begin{abstract}
 In this paper, we present a novel transformer-based architecture for end-to-end image compression. Our architecture incorporates blocks that effectively capture local dependencies between tokens, eliminating the need for positional encoding by integrating convolutional operations within the multi-head attention mechanism. We demonstrate through experiments that our proposed framework surpasses state-of-the-art CNN-based architectures in terms of the trade-off between bit-rate and distortion and achieves comparable results to transformer-based methods while maintaining lower computational complexity. 
\end{abstract}


\section{Introduction}

Transform coding is a widely employed method for compressing images and serves as the foundation for several popular coding standards like JPEG. Codecs that utilize transform coding typically consist of three components for lossy compression: transform, quantization, and entropy coding. These components have seen advancements through the application of deep neural networks and end-to-end training, as evidenced by various studies~\cite{balle2018variational,Learning_Convo,CondiProModel,NEURIPS2018_53edebc5,lee2018contextadaptive,9190935}.

Among the early works, the authors of~\cite{balle2018variational} introduced a CNN-based two-level hierarchical variational autoencoder with a hyper-prior serving as the entropy model. This architecture comprises two sets of encoders/decoders one for the generative model and another for the hyper-prior model.

Recently, transformers~\cite{vit} have demonstrated remarkable success in computer vision, including neural image compression. The authors of~\cite{Entroformer} incorporated the attention mechanism into the image compression framework by introducing self-attention in the hyper-prior model. Additionally, More sophisticated Swin block~\cite{Swin} in both the generative and hyper-prior models~\cite{zhu2022transformerbased}, utilizing shift window-based attention to confine attention to local windows. Unlike convolutional neural networks, transformers possess the ability to adapt their receptive field based on the task, with the attention mechanism's capacity to handle global context. This enhanced understanding of global information enables the capture of long-range dependencies in image compression applications.

Positional encoding holds significant importance in transformers. In the original ViT transformer~\cite{vit}, images are divided into non-overlapping patches, each mapped to a token. The standard transformer layers then process the entire token sequence simultaneously. Therefore, positional encoding plays a crucial role in preserving the sequence order and various approaches have been proposed to better model the positional information and maintain local context~\cite{attn,relativepos,ConddiPos}. In the context of image compression, positional encoding has demonstrated benefits in terms of Rate-Distortion (RD) performance, as shown in works such as~\cite{Entroformer,zhu2022transformerbased}.

Despite its advantages, employing positional encoding in transformers can increase the dimensionality of embeddings, leading to higher computational costs during training and limiting model flexibility. Recently, the authors of~\cite{cvt} demonstrated that positional encoding can be omitted in the attention module for image classification without any performance drop. They achieved this by introducing convolution in the tokenization process of patches and the self-attention block to preserve local spatial information. This combination of convolution and the attention mechanism leverages the advantages of both convolutional neural networks and transformers.

In this paper, we introduce a novel image compression framework called SwinNPE. It's built upon our proposed \textit{convolutional Swin block}, which integrates patch convolution and shift window-based attention in Swin, eliminating the requirement for positional encoding. 

We believe that this framework excels in capturing spatial contextual information more effectively. Preliminary experiments demonstrate that SwinNPE achieves comparable results to the SwinT architecture~\cite{zhu2022transformerbased} while eliminating the need for positional encoding and utilizing fewer parameters. Some of the results of this paper have been presented at~\cite{Bouzid2023}.

\section{Proposed framework}

The proposed SwinNPE uses the same architecture as in~\cite{zhu2022transformerbased}, which is shown in Figure~\ref{fig:pipeline}. Specifically, the input image $x$ is first encoded by the generative encoder $y=g_a(x)$, and the hyper-latent $z=h_a(y)$ is obtained. The quantized version of the hyper-latent $\hat{z}$ is modeled and entropy-coded with a learned factorized prior to passe through $h_s(\hat{z})$ to obtain $\mu$ and $\sigma$ which are the parameters of a factorized Gaussian distribution $P(y|\hat{z})= \mathcal{N}(\mu,\,diag(\sigma))$ to model $y$. The quantized latent $\hat{y}=Q(y-\mu)+\mu$ is finally entropy-coded (Arithmetic encoding/decoding AE/AD) and sent to $\hat{x}=g_a(\hat{y})$ to reconstruct the image $\hat{x}$. We use the classical strategy of adding uniform noise to simulate the quantization operation (Q) which makes the operation differentiable. The channel-wise autoregressive block~\cite{Learning_Convo,CondiProModel} is designed to learn the auto-regressive prior which factorizes the distribution of the latent as a product of conditional distributions incorporating prediction from the causal context of the latents~\cite{NEURIPS2018_53edebc5,lee2018contextadaptive,9190935}.

The generative and the hyper-prior encoder, $g_a$ and $h_a$, are built with the patch merge block and the convolutional Swin block. The patch merge block contains the \textit{Depth-to-Space} operation~\cite{zhu2022transformerbased} for down-sampling, a normalization layer, and a linear layer to project the input to a certain depth $C_i$. In $g_a$, the depth $C_i$ of the latent representation increases as the network gets deeper which allows for getting a more abstract representation of the image. The size of the latent representation decreases accordingly. In each stage, we down-sample the input feature by a factor of $2$.

\begin{figure*}
  \centerline{\includegraphics[width=\textwidth]{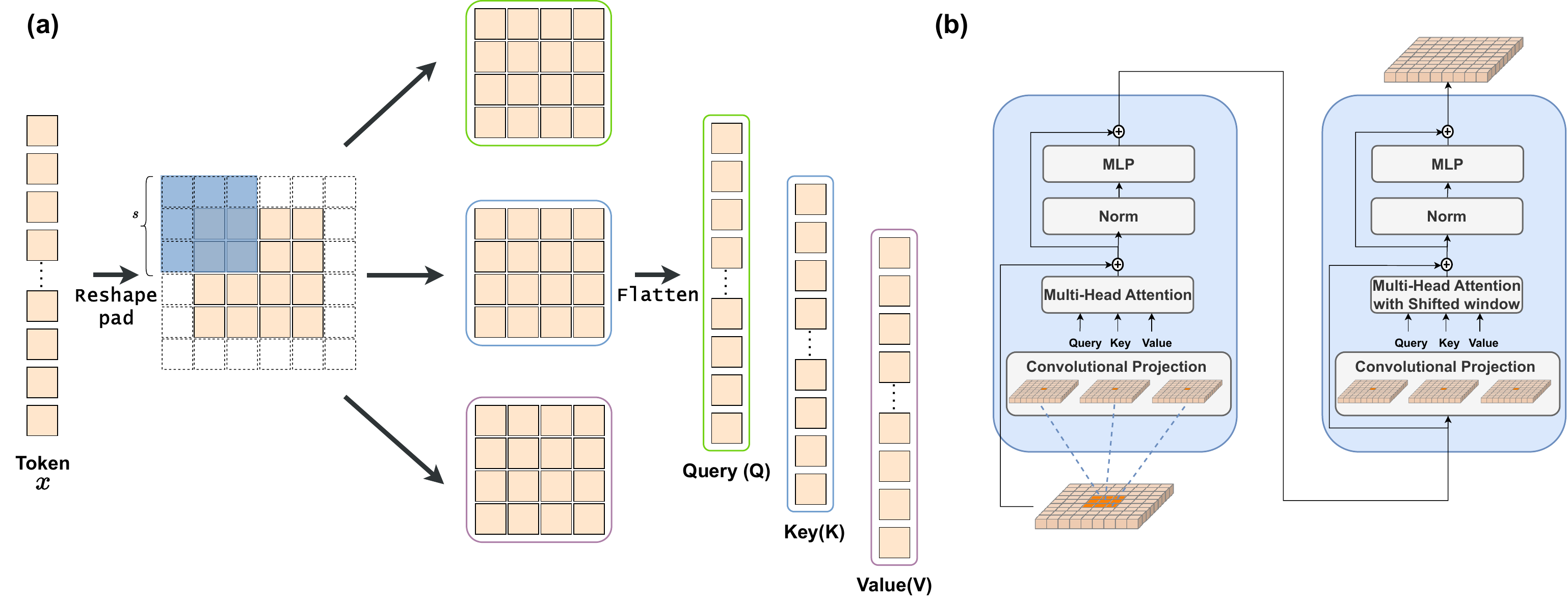}}
  \caption{(a) Convolutional Projection. (b) Convolutional Swin Block.}
  \label{fig:conv_swin_cell}
\end{figure*}

\begin{figure}[h]
    \centering  
    \subfigure(a){\includegraphics[width=0.21\textwidth]{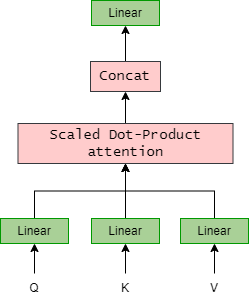}} 
    \subfigure(b){\includegraphics[width=0.21\textwidth]{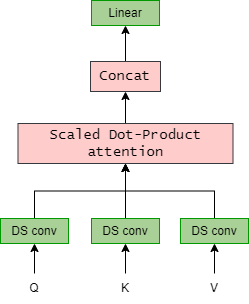}} 
    \caption{(a) The attention mechanism scheme for multi-head attention (b) The attention mechanism scheme for convolutional Swin. DS conv means depthwise separable convolution.}
    \label{fig:attention_scheme}
\end{figure}

The convolutional Swin block proposed in this work is an extension of the Swin cell~\cite{Swin}, as illustrated in Fig.~\ref{fig:attention_scheme}. Instead of employing position-wise linear projections, we utilize convolutions to project the K, Q, and V matrices within the multi-head attention block. Rather than relying on hand-crafted positional encoding, we leverage the convolution layer to capture positional information. More specifically, following~\cite{cvt}, we reshape the tokens into 2D dimensions and apply 2D-convolution and flattening operation to get tokens more sensitive to spatial context as illustrated in Fig.~\ref{fig:conv_swin_cell}:      
\begin{equation}
K,Q,V =\text{Flatten}(\text{Conv2d}(\text{Reshape2D}(x)))
\end{equation} 

To achieve parameter efficiency, we employ depth-wise separable convolution~\cite{cvt}. Specifically, the depth-wise separable convolution performs a 2D convolution independently in each feature channel. The results are then concatenated and passed through another convolution layer. This approach reduces the number of parameters and computations while enhancing representational efficiency, as it operates not only on spatial dimensions but also on the depth dimension. Importantly, it should be noted that the proposed block is not limited to convolution operations; different forms of convolution are possible~\cite{dai2017deformable,chi2020fast}, making the convolutional Swin block highly adaptable. Unlike the convolutional attention block mentioned in~\cite{cvt}, we retain the shift window structure, which facilitates cross-window connections.

The generative and hyper-prior decoders, denoted as $g_s$ and $h_s$ respectively, are constructed using the patch split block and the convolutional Swin block. In the patch split block, we reverse the merging sequence and \textit{Space-to-Depth} operation~\cite{zhu2022transformerbased} for up-sampling.

\section{Experiment and Analysis}
\subsection{Experiment configuration}

This section presents an assessment of the SwinNPE architecture and a comparison of its image compression results against state-of-the-art approaches. The SwinNPE was trained on the CLIC2020 training set for 3.3 million steps. During training, each batch consisted of eight randomly cropped images with a size of $256\times256$ pixels. 

\begin{figure}[h]
  \includegraphics[width=0.5\textwidth]{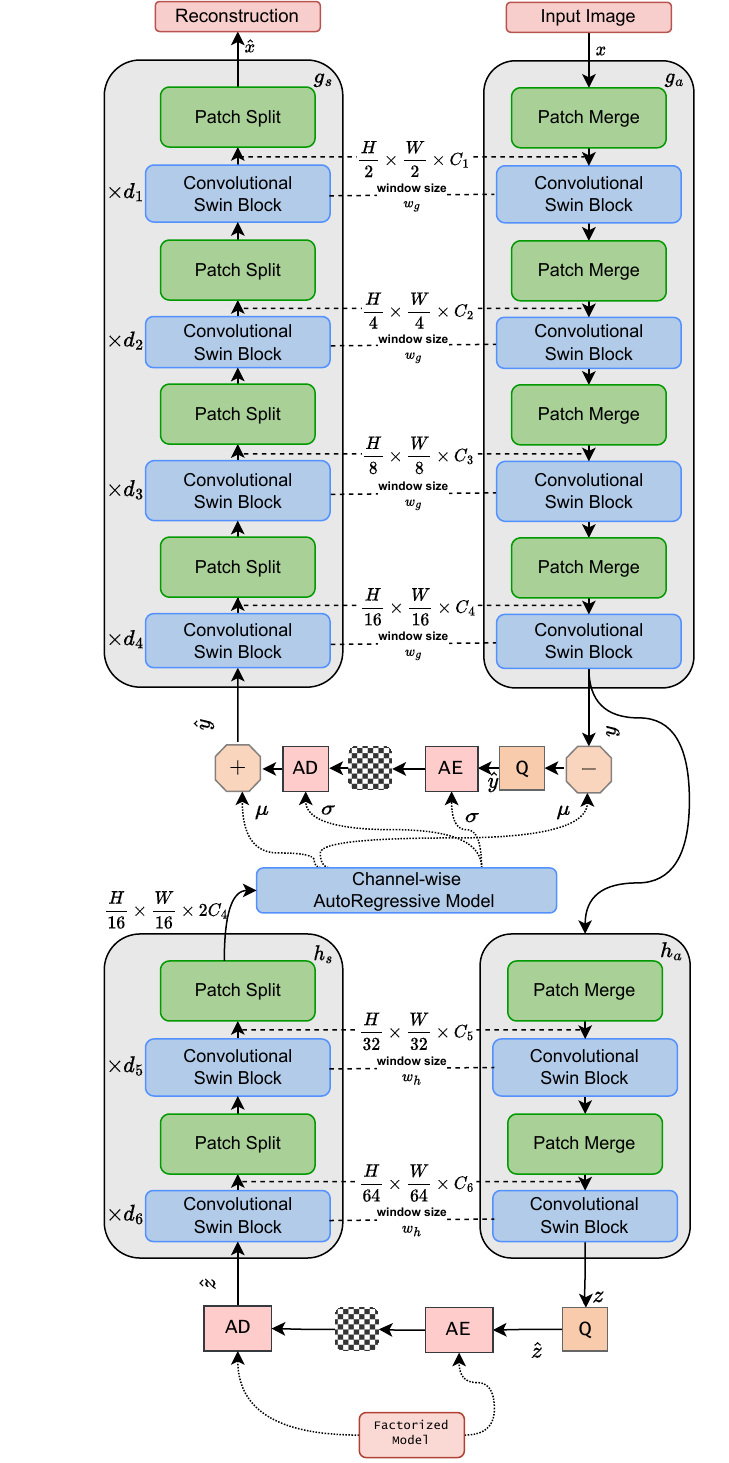}
  \caption{Network architecture of our proposed SwinNPE.}
  \label{fig:pipeline}
\end{figure}

The SwinNPE's performance was evaluated on the Kodak and JPEG-AI test dataset~\cite{Kodak,jpegai} and we center-cropped all images to multiples of 256 to avoid padding. We choose the following loss function to optimize the trade-off between the bit-rate $R$ and the quality of reconstruction $D$ which corresponds to the Mean Squared Error (MSE) in RGB color space:
\begin{equation}
L= D + \beta R,
\label{eq}
\end{equation}
with $\beta \in \{0.003, 0.001, 0.0003,0.0001\}$.

\begin{table*}[h]
\caption{Performance comparison using Bijønteguard metric~\cite{Bjntegaard2001CalculationOA} for Kodak dataset~\cite{Kodak} where $\Delta$PSNR measures the average PSNR difference and \% $\Delta$ rate the average rate saving in percentage between SwinT-CHARM~\cite{zhu2022transformerbased} (selected as the reference network) and another given network. -* GMACs of the corresponding model are not provided}
\centering

\begin{tabular}{|c|c|c|c|cc|l}
\cline{1-6}
Network &  \#Param. (M) & GMACs & Positional Encoding      & \multicolumn{2}{c|}{Bijønteguard Metric} &  \\ \cline{5-6}& & & & \multicolumn{1}{c|}{$\Delta$ PSNR} & $\% \Delta$ rate &  \\ \cline{1-6} SwinT-CHARM~\cite{zhu2022transformerbased}& 32 & 223 & Positional Relative Encoding 2D & \multicolumn{1}{c|}{0}        & 0\%      &  \\ \cline{1-6}Entroformer~\cite{Entroformer}& 142.7&        -*                & Positional Relative Encoding 2D + Diamond & \multicolumn{1}{c|}{-0.228}   & 4.33\%   &  \\ \cline{1-6}
\textbf{SwinNPE (Ours)}            & 27                                                                      & 178                    & -                                         & \multicolumn{1}{c|}{-0.311}   & 5.46\%   &  \\ \cline{1-6}
\end{tabular}

\label{tab}
\end{table*}

The schedule learning rate starts at $10^{-4}$ and the hyper-parameters of the architecture shown in  Fig.~\ref{fig:pipeline} are as follows:
$(d_1,d_2,d_3,d_4,d_5,d_6) = (2,2,6,2,5,1)$, $(w_g,w_g) = (8,8)$, $(w_h,w_h) = (4,4)$, and $(C_1,C_2,C_3,C_4,C_5,C_6) = (128,192,256,320,192,192)$. For the autoregressive model, we use the model proposed in~\cite{9190935} with 10 slices. The kernel size in all convolutional Swin blocks for depth-wise separable convolution is set to~3. 

\subsection{Analysis}

\begin{figure}[h]
    \centering
    \includegraphics[width=0.5\textwidth]{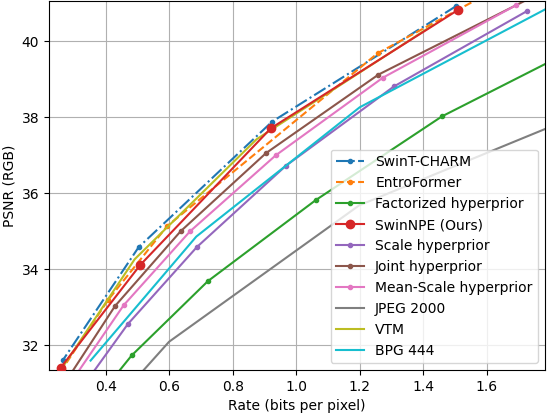}
    \caption{SwinNPE achieves nearly the same results as Entroformer~\cite{Entroformer} and SwinT-CHARM~\cite{zhu2022transformerbased} that relying on Positional encoding and better RD performance than CNNs-based methods Factorized~\cite{balle2016end}, Scale~\cite{balle2018variational}, Mean-Scale~\cite{NEURIPS2018_53edebc5}, Joint hyperprior~\cite{NEURIPS2018_53edebc5} and standard codecs on the Kodak~\cite{Kodak} image set.}
    \label{fig:curve_BD}
\end{figure}

\begin{figure}[h]
    \centering
    \includegraphics[width=0.5\textwidth]{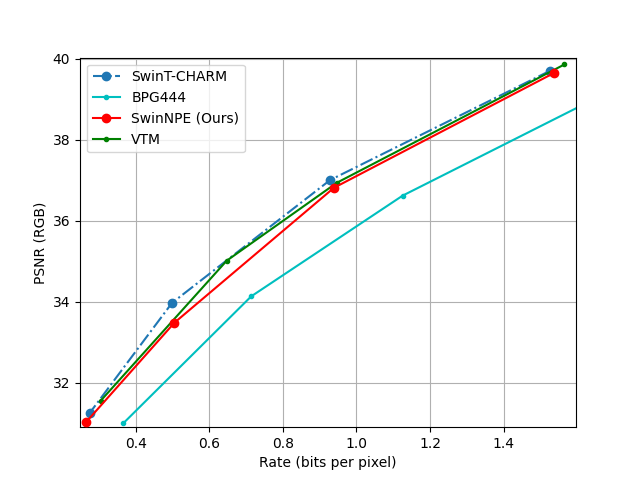}
    \caption{SwinNPE achieves nearly the same results as SwinT-CHARM~\cite{zhu2022transformerbased} and better RD performance than standard codecs on the JPEG-AI test-set~\cite{jpegai}.}
    \label{fig:curve_BD_jpegai}
\end{figure}

We compare our proposed SwinNPE model with the results of two transformers-based architectures~\cite{zhu2022transformerbased,Entroformer} and some of the most used CNN-based image compression architectures and standard codecs. The rate-distortion curves of different methods are shown in Figure~\ref{fig:curve_BD} and Figure~\ref{fig:curve_BD_jpegai} on the Kodak dataset~\cite{Kodak} and JPEG-AI test-set~\cite{jpegai} respectively. In these two figures, the PSNR and the rate shown are the average values across all images of the respective datasets. We summarize the number of parameters and GMACs of the tested transformer-based architectures in Table~\ref{tab} where we also illustrate the Bijønteguard metric~\cite{Bjntegaard2001CalculationOA} using the SwinT-CHARM as the reference for the Kodak dataset~\cite{Kodak}.

From Figure~\ref{fig:curve_BD}, we can clearly see that the SwinNPE outperforms all of the tested CNN-based architectures in terms of the bit-rate/distortion tradeoff. It is particularly interesting to notice that our proposed approach obtains almost the same results as Entroformer~\cite{Entroformer} (orange dashed line in Figure~\ref{fig:curve_BD}) with much fewer model parameters (see Table~\ref{tab}). Specifically, the saving bit-rate of SwinNPE is 5.46\% less than SwinT-CHARM (optimal saving bit-rate) which is at the same level as Entroformer with 4.33\% more bit-rate saving compare to SwinT-CHARM. We argue that it is due to the fact that the convolutional layer in the proposed convolutional Swin block can capture the local contextual information. From Figure~\ref{fig:curve_BD} and Figure  ~\ref{fig:curve_BD_jpegai}, we can see that with fewer parameters, the proposed SwinNPE has results comparable to SwinT-CHARM on both datasets. We emphasize that our proposed architecture is particularly advantageous compared to SwinT-based architecture without positional encoding~\footnote{The results are shown in the ablation studies in~\cite{zhu2022transformerbased}.} validating the advantages of combining convolutions and transformers for image compression. 

\begin{figure*}[h]
    \centering
    \includegraphics[width=0.95\textwidth]{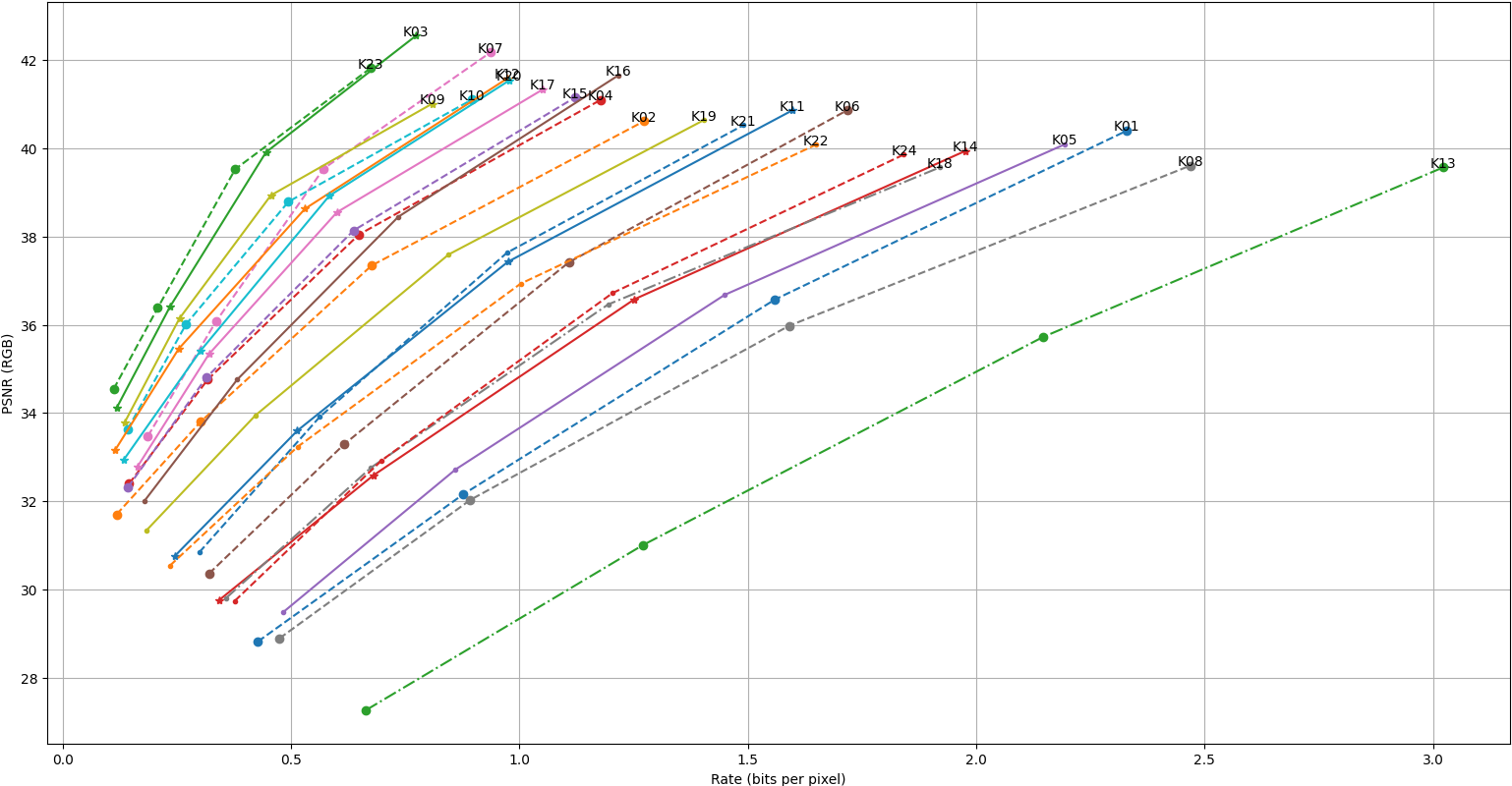}
    \caption{Rate-Distortion (RD) performance of SwinNPE on each image of Kodak\cite{Kodak} dataset.}
    \label{fig:curve_BD_details}
\end{figure*}

Figure~\ref{fig:curve_BD_details} provides the Rate-Distortion (RD) curves on each image of the Kodak dataset~\cite{Kodak}. Each curve presents the evaluation of an image with different versions of $\beta$ value (i.e. different SwinNPE models). As expected, experiments with the same $\beta$ values reveal variability in PSNR and rate values across different images. From the curves, a significant difference can be seen between images, highlighting the substantial discrepancy in results when using the same approach on different images. This observation highlights the dependence of the results on the individual characteristics of the images in the dataset.

\begin{figure*}[h]
    \centering
    \includegraphics[width=\textwidth]{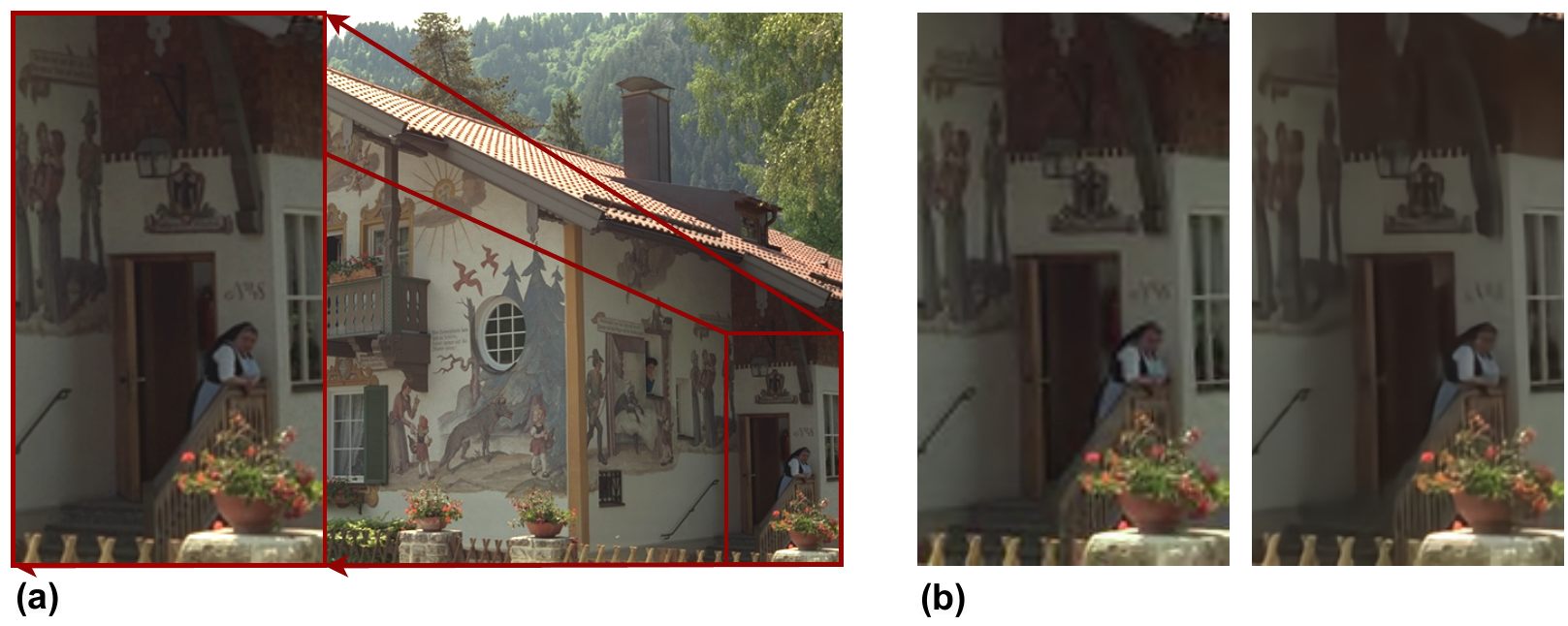}
    \caption{(a) One example (K24) of the original image from Kodak dataset \cite{Kodak}. (b) Left: image compressed with JPEG2000 (0.9257 bpp and PSNR: 30.6395 dB). Right: image compressed with the proposed SwinNPE for $\beta = 0.001$ (0.6987 bpp and PSNR: 32.9314 dB).}
    \label{fig:rendu}
\end{figure*}

We inspect the visual quality in Figure~\ref{fig:rendu} of the proposed approach with a standard codec as a reference. Figure~\ref{fig:rendu}~(a) presents a crop of one original image (K24) from the Kodak dataset~\cite{Kodak}. Figure~\ref{fig:rendu}~(b) compares the reconstructed cropped image using JPEG2000 with our method, showing that SwinNPE preserves more details even if the bit-rate is smaller than those of JPEG2000.

\section{Conclusion}
This paper introduces SwinNPE, an image compression model based on transformers that utilizes convolutional Swin blocks instead of positional encoding. SwinNPE achieves comparable performance to state-of-the-art methods while employing fewer parameters and surpassing CNN-based architectures.

The convolutional Swin block proposed in this work enables enhanced utilization of spatial context without relying on positional encoding, thereby providing increased flexibility and reducing the number of parameters.

In future research, it would be interesting to explore the utilization of diverse convolution operations and sizes within the SwinNPE model. This exploration could enable more precise modeling of complex spatial relationships and patterns, ultimately leading to improved image compression performance. 

Furthermore, integrating the convolution operation into the patch merge/split module could leverage the advantages of CNNs. The incorporation of convolutional Swin blocks in the SwinNPE model offers a promising avenue for developing efficient and effective transformer-based models for image compression.\\


\bibliographystyle{IEEES}
\bibliography{Convolutional_Transformer-Based_Image_Compression}

\end{document}